\documentclass[12pt]{article}
\usepackage{amsmath}
\usepackage{amsfonts}
\usepackage{amsthm}
\newtheorem{thm}{Theorem}[section]
\newtheorem{lem}{Lemma}[section]

\theoremstyle{remark}
\newtheorem{rem}{Remark}
\numberwithin{equation}{section}
\begin{document}

\title{Rotating Fluids with Self-Gravitation in  Bounded Domains \footnote { This paper has appeared in {\it
 Arch. Rational Mech. Anal.}
173 (2004) 345-377.} }

\author{ \\ TAO LUO \&  JOEL SMOLLER
 }
\date{}
 \maketitle

\begin{abstract}
In this paper, we study the steady solutions of Euler-Poisson
equations in  bounded domains with prescribed angular velocity.
This models a  rotating Newtonian  star consisting of a
compressible perfect fluid with given equation of state $P=e^S\rho^{\gamma}$.
 When the domain is a ball and the angular
velocity is  constant, we obtain both existence and non-existence theorems, depending on the
adiabatic gas constant $\gamma$. In addition we obtain some interesting properties of the solutions; e.g.,
 monotonicity of the radius of the star  with both  angular velocity and central
density. We also prove that the radius of a rotating spherically symmetric star, with
given constant angular velocity and constant entropy,  is
uniformly bounded independent of the central density . This is
physically striking and in sharp contrast to the case of the
nonrotating star.  For general domains and variable angular
velocities,  both an existence result for the isentropic equations of state
and non-existence result for the non-isentropic equation of state are also
obtained.
\end{abstract}

\section{Introduction}

The purpose of this paper is to understand the structure of a
rotating star. In the theory of General Relativity, for a
non-rotating star consisting  of a perfect fluid, the exterior and
interior solutions are very well understood. The exterior solution
of the star, that is, the gravitational field (space time metric)
outside the star,  is given by the well-known Schwarzschild
solution, and the interior solution, which matches the exterior
Schwarzschild solution on the boundary of the star,
 is obtained by solving the  Tolman-Oppenheimer-Volkoff
equations \cite{wen}. For a rotating star, the exterior solution
of the gravitational field is the celebrated Kerr solution
\cite{adler}. However, no corresponding interior solution of
Einstein's equation is known which matches  the Kerr solution at
the boundary of the rotating star. Actually, even in the Newtonian
case, unlike the case of the non-rotating star which has been
extensively studied by Ritter, Lane, Emden, Kelvin and
Chandrasekhar; (see \cite{ch}; and also \cite{LY} and \cite{MK} for
more recent results in the case of non-rotating star),   there have
been very few results on the rotating star if the star consists of a compressible perfect fluid.  This
paper is devoted to the investigation of the steady solutions of a
compressible perfect fluid  rotating star in the Newtonian case
with  prescribed angular velocity, in a bounded domain, and with zero
density on the boundary chosen to match the exterior of the star.

 The motion of the
 compressible perfect fluid with  self-gravitation is
modelled by the following Euler-Poisson equations;  cf \cite{ch},
\begin{equation}\label{1}
\begin{cases}
&\rho_t+{\rm div}_{ x}(\rho \bf v)=0,\\
&\rho {\bf v}_t+(\rho{\bf v}\cdot \nabla){\bf v}+\nabla P+\rho \nabla \Phi=0,\\
& S_t+{\bf v}\cdot \nabla S=0,\\
&\Delta \Phi=4\pi G \rho,
\end{cases}
\end{equation}
where $\rho$, $\bf v$, $P$, $S$ and $\Phi$ denote the density,
velocity, pressure, entropy and gravitational potential,  respectively. Here
$(t,
 x)\in \bf R^+ \times \bf R^3$ denotes the time and space
variables and $ x=(x_1, x_2, x_3)$. For  simplicity, we assume the
pressure satisfies the following equation of state
\begin{equation}\label{P}
P=P(\rho,\ S)=e^S\rho^{\gamma}, \end{equation} where $\gamma>1$ is the
adiabatic  exponent.

 Suppose the star rotates about the
$x_3$-axis;  we are interested in finding an axi-symmetric solution $(\rho,\
{\bf v},\ S,\  \Phi)(x,\  t)= (\rho,\  {\bf v},\ S,\  \Phi)(\eta(x), x_3, t)$ of
(\ref {1}) with prescribed time-independent angular velocity
$\Omega(\eta)$, where $$ \eta=\eta(x)=\sqrt{x_1^2+x_2^2}.$$ In this case the
velocity field is given by ${\bf v}=(-x_2\Omega(\eta),
x_1\Omega(\eta), 0)$, and thus ${\rm div}_x(\rho {\bf v})=0$ and
${\bf v}\cdot \nabla S=0$. By
$(1.1)_1$ and $(1.1)_3$, $\rho_t=0$ and $S_t=0$. Thus, the solution $(\rho,
{\bf v},\ S,\ \Phi)$
of (1.1) is time-independent and satisfies the following system of equations
\begin{equation}\label{2}
\begin{cases}
&\nabla P(\rho,\ S)+\rho \nabla \Phi-\rho\nabla J(\eta)=0,\\
&\Delta \Phi=4\pi G \rho,
\end{cases}
\end{equation}
where
\begin{equation}\label{J}
J(\eta)=\int_0^{\eta} s \Omega^2(s)ds,
\end{equation}
 and the entropy $S$ is a given bounded $C^1$ function.

In \cite{AB}, for an isentropic fluid, i.e., the entropy $S=constant$ (without loss
of generality, S is assumed to be zero in this case),  Auchmuty and Beals considered   system (\ref{2}) in
all of $R^{3}$, with  prescribed total mass $M$, and
gravitational potential $\Phi$ given by
$$\Phi(x)=-G\int_{R^3} \frac{\rho(y)}{|x-y|}dy=:-B\rho(x).$$
 The
problem then reduces to finding  the solution of the following
equation,
\begin{equation}\label{L}
\nabla(\frac{\gamma\rho^{\gamma-1}}{\gamma-1}-B\rho-J(\eta))(x)=0,
\end{equation}
where $\rho>0$. They formulated this as a variational problem;
namely, minimize
  \begin{equation}\label{E}
  E(\rho)=\int_{R^3}
  (\frac{\rho^{\gamma}}{\gamma-1}-\frac{1}{2}\rho\cdot B\rho-\rho
  J(\eta))(x)dx,
  \end{equation}
  in the class
  \begin{equation}\label{W}
  W_M=\{\rho\ge 0:\int_{R^3}\rho(x)dx=M\}.
  \end{equation}
By assuming that  the angular velocity $\Omega$ satisfies the following
decay properties
\begin{align}\label{J1}
& J(+\infty)<+\infty, J\in C^1[0, +\infty),\notag\\
&\eta(J(+\infty)-J(\eta))\to 0,\ as\  \eta\to +\infty,
\end{align}
and that the adiabatic exponent $\gamma$  satisfies
\begin{equation}\label{4/3}
 \gamma>4/3,
 \end{equation}
Auchmuty and Beals proved the existence of a minimizer of the
functional $E(\rho)$ in the class of functions  (\ref{W}).
Moreover, this minimizer  has compact support and satisfies
equation (\ref{L}) wherever it is positive. The shape of the free
boundary which separates the vacuum and fluid was investigated in
\cite{CF} for the Auchmuty-Beals solutions.  The
case  of an isentropic
 uniformly rotating star,
(i.e., the
angular velocity $\Omega$ is  constant),  was discussed by Li
 in (\cite{Li1}); he proved the existence of a minimizer of the functional (1.6) in
 the class $W_M$ ,  under the  assumption  $\gamma>4/3$. The diameter of the support
 of the density $\rho$ was studied in \cite{Li2} for the solution obtained
 in \cite{Li1}.   In the proof of the above
results,  the prescribed total mass serves as a constraint on
these variational problems, and without this constraint, it is not
clear that the minimizer of $E(\rho)$  exists. In \cite{AB}, the
angular velocity is prescribed in the entire space $R^3$ (even in
the vacuum region), and is assumed to satisfy the decay property
(\ref{J1}).
 In  \cite{AB} and \cite {Li1}, $\gamma$ is required to be greater
 than $4/3$, so a natural question one can ask is  what happens
 when $1<\gamma\le 4/3$ ? Another issue is that all of  the above mentioned results
are  for  isentropic fluids, so one can also ask what happens if the fluid
is non-isentropic.  The purpose of this paper is to address
 these, and other issues.

\section{Statement of Results.}

 We are interested in the solution of (\ref{2}) in a bounded domain $ D
$ in $R^3$, where $\rho( x)>0$ for $ x\in D$ and $\rho( x)=0$ if $
x\in
\partial D$. From the first equation in system (1.3), we have
\begin{equation}\label{meng1}
\frac{1}{\rho}\nabla P=\nabla(J-\Phi)\end{equation} for $x\in D$.
Thus, by the second equation of (1.3), we obtain
\begin{equation}\label{meng2}
div(\frac{1}{\rho}\nabla P)=\Delta (J-\Phi)=\Delta J-4\pi
G\rho\end{equation} for $x\in D$.  Set
\begin{equation}\label{n}
w=\frac{\gamma}{\gamma-1}(e^{S/\gamma}\rho)^{\gamma-1}.\end{equation}
Then it is easy to verify, using (1.2), that
\begin{equation}\label{meng3}
\frac{1}{\rho}\nabla P=e^{S/\gamma}\nabla w.
\end{equation}
Substituting this in (\ref{meng2}),  we obtain   the
following elliptic equation
\begin{equation}\label{3}
div (e^{\alpha S}\nabla w)+Ke^{-\alpha S} w^q-2\Omega(\eta)( \Omega(\eta)+\eta \Omega'(\eta))=0,
\end{equation}
 where
\begin{equation}\label{1.12*}
q=\frac{1}{\gamma-1},\ \ \alpha =\frac{1}{\gamma}, \end{equation}
$K=4\pi G\left(\frac{\gamma-1}{\gamma}\right)^{\frac{1}{\gamma-1}},$
 and for
simplicity, we can normalize $K$ to make $K=1$.  We  seek
solutions of (\ref{3}) satisfying
\begin{equation}\label{4}
\rho(x)>0,\  x\in D,\  \rho( x)=0, \  x\in \partial D,
\end{equation}
or equivalently
\begin{equation}\label{n2}
w(x)>0,\  x\in D,\  w(x)=0, \  x\in \partial D.
\end{equation}
In this paper, we only consider the case when
$$1<\gamma<2,$$
because when $\gamma>2$, $0<q<1$, and equation (\ref{3}) becomes sublinear,
and this situation was studied completely  in \cite{SW}. Moreover,  equation (\ref{3}) is linear  if $\gamma=2$,
and there is a complete theory for linear elliptic equations (cf. \cite{GT}).

We first consider the   case when the angular velocity
$\Omega(\eta)=\Omega=constant$, the entropy $S(x)$ is spherically symmetric (
i.e., $S(x)=S(r),\ r=|x|=\sqrt{x_1^2+x_2^2+x_3^2}$),
and the domain $D$
is a ball $B_R(0)$.   In this case,
we look for  the spherically
symmetric solutions of  problem (\ref{3}) and (\ref{n2}), i.e. ,
$w(x)=w(r), \  r=|x|=\sqrt{x_1^2+x_2^2+x_3^2}$. Then $w(r)$
satisfies the following equation
\begin{equation}\label{100}
w''(r)+(\frac{2}{r}+\alpha S'(r))w'(r)+e^{-\alpha S(r)}(e^{-\alpha S(r)}w^q-\sigma)=0,
\end{equation}
and boundary conditions
\begin{equation}\label{101}
w'(0)=0,\  w(R)=0,
\end{equation}
where we have set
\begin{equation}\label{1.16*}
\sigma=2\Omega^2.
\end{equation}
We
want to use the ''time-map'' method used by Smoller and Wasserman
(cf \cite{SW}). For this purpose, we consider the problem
(\ref{100}) with initial data
\begin{equation}\label{102}
w(0)=p>0,\  w'(0)=0,
\end{equation}
 where $p$
is a free parameter,  and let  $w(r, p, \sigma)$ be  the solution of
this problem. We define the '' time-map'' $p\to R(p,\ \sigma)$, by
\begin{equation}\label{smoller21} R(p, \sigma)=\inf\{R|\ R>0,\  w(r, p, \sigma)>0  {~\rm~} if \  0\le r< R, \ and
\ w(R, p, \sigma)=0\}, \end{equation} so $R(p,\ \sigma)$ is the first ''time''
at which $w$ is 0 (we will write $R(p,\ \sigma)=\infty$ if $w(r,\ p,\ \sigma)>0$ for all $r\ge 0$. ) Thus $R(p, \sigma)$ is the radius of a rotating
star with given central density (cf. (\ref{n}))
$$\rho(0)=(\frac{\gamma-1}{\gamma}p)^{\frac{1}{\gamma-1}}e^{-S(0)/\gamma}.$$
 and
angular velocity $$\Omega=\sqrt{\sigma/2}.$$  In order to state our first theorem,
we introduce the following notation. We   set
\begin{equation}\label{meng6}
\bar S=\sup_{r\ge 0} S(r), \underline S=\inf_{r\ge 0} S(r).
\end{equation}
We assume
\begin{equation}\label{meng7}
-\infty<\underline S\le \bar S<\infty.
\end{equation}
We define the constant $b$ by
\begin{equation}\label{b}
b=\frac{(10+22q)e^{-\frac{\alpha\bar
S}{q}}\sigma^{\frac{q-1}{q}}}{(1+2e^{\alpha \bar
S})[10(1+q)]^{\frac{1}{q}}(5-q)^{1-\frac{1}{q}}} .
\end{equation}
Obviously $b>0$ if $1<q<5$.  Our  first  result
is the following theorem.
\begin{thm}\label{thm2.1} Assume that $D$ is a ball in ${\bf R}^3$ and that $\Omega$ is a
nonzero constant.\\
1) If~  $1<\gamma\le 6/5$ and $S'(r)\ge 0 $ for $r\ge 0$, then
$$w(r, p, \sigma)>0,$$  for every $r\ge 0$ and $p>0$ . \\

\noindent 2) If~   $4/3<\gamma<2$, then
 \begin{equation}
R(p, \sigma)<+\infty,
\end{equation}
for sufficiently large $p$,  provided the entropy satisfies the
following conditions
\begin{equation}\label{s'}
\alpha S'(r)e^{\alpha S}z'>-\frac{b}{2}z,
\end{equation}
for $0<r<\pi/\sqrt{b},$ where
\begin{equation}\label{z3}
z(r)=\frac{\sin(\sqrt{b}r)}{\sqrt{b}r},
\end{equation}
and $\sup_{0\le r<1+\frac{\pi}{\sqrt b}} |S'(r)|$ is sufficiently small. \\

\noindent 3) If~   $6/5<\gamma<2$, assume that the conditions in
2) hold, and  assume too that the entropy satisfies the following
condition
\begin{equation}\label{z38}
S'(r)\le 0, {\rm~ for~} 0\le r\le \frac{\pi}{\sqrt b},
\end{equation}
where $b$ is given by (\ref{b}).  Then
 \begin{equation}
R(p, \sigma)<+\infty,
\end{equation}
for sufficiently large $p$. \\

\noindent 4) For the solutions $w(r,\ p,\ \sigma)=:w(r)$ in 2) and
3), we have the following estimates on the  mass
$M(r)=\int_0^r4\pi \tau^2\rho(\tau)d\tau,$  and the average
density  $\bar \rho(r)=\frac{1}{Vol B_r(0)} M(r) $ in the ball
$B_r(0)$, (here $\rho=
\left[\frac{\gamma-1}{\gamma}w\right]^{1/(\gamma-1)}$ is the
density (cf. (2.3)) and $Vol\ B_r(0)=4\pi r^3/3$):
\begin{equation}\label{mass}
M(r)\ge \frac{2r^3\Omega^2}{3G},
\end{equation}
\begin{equation}\label{adensity}
\bar \rho (r)\ge \frac{\Omega^2}{2\pi G},
\end{equation}
for $r\le R(p,\ \sigma)$, where $G$ is the Newtonian gravitational
constant.
\end{thm}
\begin{rem}
The conditions imposed on the entropy $S(r)$ in Theorem 2.1 are
automatically satisfied in the isentropic case, i.e.,
$S(r)=constant$.
\end{rem}
\begin{rem} The existence of positive solutions for the equation of isentropic {\it non-rotating} star is standard
(cf. [5], [12] and [15]).  In [8], for the case of a {\it non-rotating} star $(\Omega=0)$,
the authors first consider non-isentropic equations of state, and an existence theorem is proved for a bounded
domain $D$, under the assumption that the entropy function $S$ satisfies, for some constant $a>0$,
$$\int_{D}|\nabla \phi|^2e^{-S/\gamma}dx+\frac{1}{\gamma}\int_{D}\Delta S\phi^2e^{-S/\gamma}dx\ge a \int_D \phi^2e^{-S/\gamma}dx,
$$ for all test functions $\phi\in H^1_0(D)$. Our condition (2.18) is an {\it explicit} condition on $S$ and does
not involve the second derivative of $S$,  for the case when
domain $D$ is a ball.
\end{rem}
The proof this theorem, as well as the other theorems stated in
this section will be given in sections 3-5.

In the case of isentropic fluids (i.e., $S=constant$),  we can
obtain further results on the qualitative properties of the
solutions if  the angular velocity $\Omega(\eta)=\Omega=costant$
and the domain $D$ is a ball. Without loss of generality, we may
assume
\begin{equation}\label{meng9}
S=0.
\end{equation}
   In this case, it follows from the celebrated
Gidas, Ni and Nirenberg result (\cite{Ni}) that  positive solutions to
(\ref{3}) and (\ref {4}) {\it must}  be spherically symmetric.
Substituting $S=0$ in (\ref{100}), we obtain  the
 following equation:
\begin{equation}\label{5}
w''(r)+\frac{2}{r}w'(r)+w^q-\sigma=0.
\end{equation}
 We consider the problem
(\ref{5}) with initial data
\begin{equation}
w(0)=p>0,\  w'(0)=0.
\end{equation}
We will again use $w(r, p, \sigma)$ to denote the solution of the
above problem.
   The following theorem
gives some physically interesting properties of $R(p, \sigma)$, the radius
of the star with central density (cf. (2.3) with $S=0$)
$$
\rho(0)=(\frac{\gamma-1}{\gamma}p)^{\frac{1}{\gamma-1}}.$$

\begin{thm}\label{thm4} Assume that $D$ is a ball in ${\bf R}^3$ , $\Omega$ is a
nonzero constant and the entropy $S$ is constant (we set $S=0$ for convenience). Then
the following statements hold. \\
\noindent 1) If~   $6/5<\gamma<2$, there exists  a constant
$p_0> 0$ depending
only on $\gamma$ and $\sigma$ such that \\
 \begin{equation}
R(p, \sigma)<+\infty, {~\rm if~ and~ only~ if~} p\ge p_0.
\end{equation}
Moreover, we can estimate $p_0$ from below,
\begin{equation}\label{low}
p_0\ge
\left[\frac{5\sigma(q+1)}{5-q}\right]^{1/q}=\left[\frac{10\gamma\Omega^2}{5\gamma-6}\right]^{\gamma-1}.
\end{equation}
2) If~  $6/5<\gamma<2$ ,
\begin{equation}\label{sig1}
 R(p, \sigma_1)\ge R(p, \sigma_2),\end{equation}
 provided\  $\sigma_1>\sigma_2>0$,
and
\begin{equation}\label{sig2}
R(p_1, \sigma)\ge R(p_2, \sigma),
\end{equation}\label{sig3}
if\  $p_2> p_1\ge p_0$.
\end{thm}
As a corollary of this result, we have the following theorem,
in which $p_0$ is the positive constant given in Theorem \ref{thm4}.

\begin{thm}\label{thmx} Assume that $D$ is a ball in ${\bf R}^3$,  $\Omega$ is a
nonzero constant and the entropy $S$ is constant (set $S=0$ for convenience).
  If~ $6/5<\gamma<2$,  then there exists a positive constant
$C$ depending only on $p_0$ and $\Omega$ such that
\begin{equation}\label{sig4}
R(p, \sigma)\le C p^{\frac{\gamma-2}{2(\gamma-1)}}\le
Cp_0^{\frac{\gamma-2}{2(\gamma-1)}},
\end{equation}
for $p\ge p_0$.
In particular, this implies
\begin{equation}\label{sig5}
R(p, \sigma)\to 0, \qquad {\rm as~} p\to \infty.
\end{equation}
\end{thm}
We now discuss the physical meaning of the above
two theorems.\\

{\it Physical meaning of Theorem \ref{thm4}}

 Part 1) means, for $6/5<\gamma<2$ and $\sigma=2\Omega^2>0$
,  there exits
 a critical central density,  determined by $p_0$,  depending  only on $\gamma$ and  the angular velocity
 $\Omega$, such that the radius of
star is finite if
 the central density is greater than or equal to this critical central density;  otherwise the radius of the star
 is infinite. This is in sharp contrast to a non-rotating star, for which in this range of $\gamma$ the radius is finite no matter how
 small the central density
 is (cf. \cite{ch}). The reason for this is that for the rotating star, the central density must be large
 enough to provide sufficient gravitational attraction to balance the centrifugal force due to
rotation.
    The physical meaning of part 2)  is that, for a compressible fluid the radius of the star  increases
with increasing angular velocity for fixed central density,
while the radius of star  decreases with increasing central
density for the fixed angular velocity.\\

{\it Physical meaning of Theorem \ref{thmx}}

If $6/5<\gamma<2$, in order to have the radius of the rotating
star $R(p, \sigma)$  to be finite, $p$ has to be greater than or
equal to $p_0$,  and the constant $p_0$ is completely determined
by  $\gamma$ and the angular velocity $\Omega$, as stated in
Theorem 2.1. Thus,  (\ref{sig4}) in Theorem \ref{thmx} shows that
there exists a (finite)  upper bound on the radius of a rotating
star, which is determined only by $\gamma$ and $\Omega$, and  is
independent of central density,  for the radius of the rotating
star to be finite.  This is physically striking and completely
different from non-rotating star, for which the radius can be
arbitrarily large. Actually, for the non-rotating star, the
central density can also be arbitrarily small (cf. \cite{ch}).
Moreover, as we will show later, the radius of a non-rotating star
is proportional to $(\rho(0))^{\frac{\gamma-2}{2}}$, where
$\rho(0)$ is the central density. Since $\gamma<2$, the radius of
a non-rotating  star can be very large if the central density is
very small. However, for the rotating star, the central density
cannot be arbitrarily small, as indicated in part 1) of Theorem
\ref{thm4}; cf. (2.27). Moreover, (\ref{sig5}) shows that the
radius of a rotating star tends to zero if the central density
tends to infinity.

We now turn to the case of general bounded domain $D\subset {\bf
R}^3$ and variable angular velocity $\Omega(\eta)$.  Without loss
of generality, we may assume that $0\in D$. For the existence of
positive solutions to  the boundary value problem (\ref{3}) and
(\ref{4}), we assume that the angular velocity
$\Omega(\eta)$ is $C^1$ as a function of $\eta$ and satisfies the following condition:\\

\noindent {\bf Condition A}:
 Let $B_{R_1}(0)=\{x|\ |x|<R_1\}$ be the larggest ball contained
in $D$ which is centered at origin. For $ x\in D- B_{R_1}(0)$,  assume
\begin{equation}\label{9}
\Omega (\eta)[\eta\Omega(\eta)]'\le 0,
\end{equation}
and is not identically zero; here, as before,
$\eta=\eta(
x)=\sqrt{x_1^2+x_2^2},$\  and\  $''$ prime $''$ denotes $\frac{d}{d\eta}$.  \\
\begin{rem}
Notice that (\ref{9}) is only required for $x\in D-B_{R_1}(0)$.
Therefore the angular velocity $\Omega(\eta)$ can be any $C^1$
function when $D$ is a ball.
\end{rem}

The following theorem is an existence theorem for a general domain $D$ and
variable angular velocity.
\begin{thm}\label{thm5}
Assume  $6/5<\gamma<2$, the entropy $S$ is constant (set $S=0$ for convenience),
 $\Omega (\eta(x))\in C^{1+\delta}(D)\cap C(\bar D)$ for
some $\delta$,  $0<\delta<1$ with $\partial D$ smooth. Then there exists a
constant $R_2>0$ depending only on $\Omega$ and $\gamma$ such that
if $D\subset B_{R_2}(0)$ and condition A holds,  there exists
a positive solution to (\ref{3}) and (\ref{4}). Moreover, for this
solution , the central density satisfies :
\begin{equation}\label{10}
\rho(0)\ge (\frac{\gamma-1}{\gamma})^{1/(\gamma-1)}\cdot
\frac{5\gamma}{5\gamma-6}\cdot \beta,
\end{equation}
where $\beta=\max_{ x\in \bar D}|2\Omega (\eta)(\Omega
(\eta)+\eta\Omega'(\eta)|$.
\end{thm}

Notice that as in Theorem 1.3, the size of the domain $D$ is
uniformly bounded,
where the bound depends only on $\gamma$ and $\Omega$.\\

  Let $w(x)$ be the
positive solution given in Theorem \ref{thm5}.  Notice that since
$S=0$  (2.3) implies
$\rho=(\frac{\gamma-1}{\gamma}w)^{\frac{1}{\gamma-1}}$. We can
obtain the potential function by virtue of (1.10) and (1.13).
Actually, when $S=0$, (2.4) and (2.5) imply $\nabla(w+\Phi-J)=0$.
So we  set $\Phi(x)=(C-w+J(\eta))(x)$ for $x\in D$, where $C$ is
any constant. Then $(\rho, \Phi)(x)$  is a positive solution to
the system (1.3) in the domain $D$. System (1.3) is derived by
assuming   that $\rho$ is axi-symmetric. In fact, we have the
following self consistency result.
\begin{thm}\label{new}
Suppose the conditions in Theorem 1.4 hold. If the domain $D$ is
axi-symmetric about $x_3$-axis, then the solution $w(x)$ in
Theorem \ref{thm5} is also axi-symmetric about $x_3$-axis.
\end{thm}

Our final result is  the following
 nonexistence theorem;  here $\nu (x)$ denotes the unit outer
 normal vector on the boundary,  and $\eta=\eta( x)=\sqrt{x_1^2+x_2^2};$
 \begin{thm}\label{th1}
Suppose  $\partial D\in C^1$ and  the domain $D$ is bounded and star-shaped,
i.e. $ x\cdot \nu ( x)\ge 0 $ for $x\in \partial D$, and the entropy $S$ satisfies the condition
\begin{equation}\label{entropy}
x\cdot \nabla S(x)\ge 0,\qquad {for~} x\in D.
\end{equation}
 If $\gamma
\le 6/5$ (resp. $\gamma<6/5$) and \begin{equation}\label{ll}
 \eta A'(\eta)+\frac{5A(\eta)}{2}> 0\ (resp.\  \ge 0)
\end{equation}
 for $ x\in D$, where $A(\eta)=2\Omega(\eta)(\eta\Omega(\eta))'$,  then there is
no positive solution to  (\ref{3}) and (\ref{4}).
\end{thm}
\begin{rem}
Condition (\ref{ll}) is trivially satisfied when $\Omega$ is a
 constant.
\end{rem}
\begin{rem} In [8], for the case of a {\it non-rotating} star $(\Omega=0)$, the authors prove a non-existence
theorem for $1<\gamma<6/5$, under the assumption that the entropy $S(x)$ satisfies the following three
conditions, for all $x\in D$:
$$x\cdot \nabla S\ge 0,\  \Delta S\ge 0\ {and~ } \nabla(\Delta S)\cdot x\ge 0. $$
In Theorem 2.6, we only require $x\cdot \nabla S\ge 0$ for all $x\in D$.
\end{rem}

We now make some remarks about the above theorems and their proofs.
In the theory of second order elliptic equations,
 the existence of a positive solution to
 the boundary value problem
 \begin{equation}\label{elliptic}
 \begin{cases}
& \Delta w+f(x, w)=0, x\in D,\\ & w|_{\partial D}=0,
\end{cases}
\end{equation}
 where $D\in R^n$ is a bounded open set,  has been extensively
studied either by the Min-Max method of the Mountain Pass Lemma of
Rabinowitz \cite{Robinowitz} or the Topological Degree Method
(Leray-Schauder degree) by P. L. Lions \cite{Lions}, both under the
condition  $f(x, 0)\ge 0$.  If $f(x, 0)<0$, the difficulty in
solving the boundary value problem (\ref{elliptic}) is that the
Harnack inequality is not applicable, and thus the above
variational methods (Mountain Pass Lemma),  and the
Topological Degree Method do not work in this case.
The first existence result of a positive solution  for the case
 $f(x,0)<0$ was given in (\cite{SW}) by
using the time-map method when $D$ is a ball and $f(x, w)=f(w)$
(i.e., $f$ does not depend on $x$ explicitly), this corresponds to
the case when both the entropy $S$ and the angular velocity
$\Omega$ are constant in equation (\ref{3}). For this type of
equation,  further analysis was given in \cite{castro}.  The
existence result of a spherically symmetric solution to (\ref{3})
and (\ref{n2}) generalizes the above mentioned results to the
non-isentropic case.

 If $D$ is a general domain,  \cite{SW} contains some existence results for the case
$f(x,w)=f(w)$ and $f$ is sublinear in $w$. To the best of our
knowledge, Theorem \ref{thm5}  is the first result of existence of
positive solutions for the case that $f(x, w)$ is superlinear in
$w$ and the domain $D$ is different from a ball.

The rest of this paper is organized as follows. In Section 3, we
prove Theorem 2.1. Theorems 2.2 and 2.3 are proved in Section 4.
Theorems 2.4 - 2.6 are proved in Section 5. Further discussions
are given in Section 6.

\section{Proof of Theorem 2.1.  }
In this section, we shall prove Theorem 2.1.  We define
\begin{equation}\label{f(w)}
f(r,\ w)=\begin{cases}& e^{-\alpha S(r)}w^q-\sigma, \qquad if~ w\ge 0,\\
                  & -\sigma, \qquad \qquad if~ w<0,
\end{cases}
\end{equation}
and
\begin{equation}\label{F(w)}
F(r,\ w)=\int_0^{w} f(r,\ z)dz.
\end{equation}

As in Section 2, we use the notation $w(r, p, \sigma)$  to denote
the solution of the problem
\begin{equation}\label{eq3.3}
w''(r)+(\frac{2}{r}+\alpha S'(r))w'(r)+e^{-\alpha S}f(r, w)=0,
\end{equation}
and
\begin{equation}\label{2.4}
w'(0)=0, w(0)=p>0.
\end{equation}
  Let
$R(p,\ \sigma)$ be as in (\ref{smoller21}); i.e., $R(p,\ \sigma)$
 the first point at which $w$ is 0.

 Motivated by the celebrated
Pohozaev identity (\cite{Po}),  we define the function $G(r)$ by
\begin{equation}\label{2.g}
G(r)=r^3\left(e^{\alpha S(r)}\frac{{w'}^2}{2}+F(r,\
w)\right)+\frac{1}{2}r^2e^{\alpha S} ww'(r)
+\frac{\alpha}{2}\int_0^r t^3 e^{\alpha S}S'w'^2dt,\ r\ge 0.
\end{equation}
It is easy to verify that
\begin{equation}\label{2.5}
G'(r)=3r^2(F(r,\ w)-\frac{f(r,\ w)w}{6})+r^3g(r,\ w),
\end{equation}
where
\begin{equation}\label{g}
g(r, w)=\int_0^w \frac{\partial f(r,\  z)}{\partial r}dz
=\begin{cases}& -\frac{\alpha S'(r) e^{-\alpha S}w^{q+1}}{q+1},
\qquad w\ge 0, \\
& 0, \qquad \qquad w<0;\end{cases}
\end{equation}
thus
\begin{equation}\label{2.6}
G(r)=\int_0^r \{3t^2[F(t, w)-\frac{f(t,\  w)w}{6}]+t^3 g(t, \
w)\}dt,\  r\ge 0.
\end{equation}
We now prove that \\
{\it If $1<\gamma\le 6/5$ ( $q\ge 5$) and  $S'(r)\ge 0$ for  $r\ge
0$, then
 \begin{equation}\label{eeq} R(p,
\sigma)=+\infty \end{equation} for any $p>0$ and $\sigma> 0.$ } \\
The proof is  by contradiction. Suppose $R(p, \sigma)<+\infty$,
and let $\bar R=R(p, \sigma))$. Thus
\begin{equation}\label{eqq1}
w(\bar R)=0, w(r)>0\ {\rm for~} 0\le r<\bar R. \end{equation} By
the definition of $G(r)$  (3.5),  we have $G(\bar R)=\frac{{\bar
R}^3e^{\alpha S(\bar R)} (w'(\bar
R)^2}{2}+\frac{\alpha}{2}\int_0^{\bar R} r^3 e^{\alpha
S}S'(r)w'^2dr$. Applying this to (3.8), we obtain, by virtue of
(\ref{f(w)}) and (\ref{F(w)}),
\begin{align}\label{eqq2}
&\frac{{\bar R}^3e^{\alpha S(\bar R)} (w'(\bar R))^2}{2}+\frac{\alpha}{2}\int_0^{\bar R} r^3 e^{\alpha S}S'(r)w'^2dr\notag\\
&=\int_0^{\bar R} 3r^2\left\{[\frac{1}{q+1}-\frac{1}{6})e^{-\alpha
S}w^{q+1}-\frac{5}{6}\sigma w]-
r^3\frac{\alpha}{q+1}S'(r)e^{-\alpha S}w^{q+1} \right\}dr.
\end{align}
Since $\frac{1}{q+1}-\frac{1}{6}\le 0$ for $q\ge 5$ and $S'(r)\ge
0$, (\ref{eqq2}) gives a contradiction. Thus (\ref{eeq}) holds so
Part 1) in Theorem 2.1 is proved.

 Next, we give the proof of parts 2) and 3) of Theorem 2.1.
 First, it follows from (\ref{eq3.3}) that
\begin{equation}\label{w'}
(r^2e^{\alpha S}w')'+r^2 f(r,\ w)=0.
\end{equation}
Thus from (\ref{w'}) and (2.4), we obtain
\begin{equation}\label{2.11}
w'(r)=-\frac{e^{-\alpha S(r)}}{r^2}\int_0^r t^2f(t, w(t))dt, \
r\ge 0.
\end{equation}

For fixed $p>0$,  since $w(0)=p$, by (\ref{f(w)}),  (\ref{F(w)})
and (\ref{2.11}), we have $w'(r)<0$ for small $r$, $r>0$,  if $p$
is sufficiently large. We define $r_1$ to be the point such that
\begin{equation}\label{t2}
w(r_1)= 2p/3, \ w(r)>2p/3\ for\ 0<r<r_1.
\end{equation}
Then we have the following lemma which  estimates $r_1$ and $G(r_1)$
 in terms of $p$.
\begin{lem}\label{lem 2.1}  For sufficiently large $p$ , if $1<q<5$, ($6/5<\gamma<2$),
then there are positive constants $c_1$, $c_2$ and $c_3$
independent of $p$ such that
\begin{equation}\label{r1}
c_1p^{-(q-1)/2}\le r_1\le c_2p^{-(q-1)/2},
\end{equation}
  and
\begin{equation}\label{2.1x}
G(r_1)\ge c_3 p^{(5-q)/2},
\end{equation}
provided $\sup_{0\le r\le  r_1}|S'(r)|$ is sufficiently small.
\end{lem}
\begin{proof}
By the definition of $r_1$ (cf. (\ref{t2}), we have
\begin{equation}\label{2.12}
2p/3\le w(r)\le p, \qquad 0\le r\le r_1.
\end{equation}
Thus, from (\ref{f(w)}), (\ref{2.11}) and (\ref{2.12}),
 \begin{align}\label{t1}
 &2p/3=w(r_1)=p-\int_0^{r_1}\frac{e^{-\alpha S(r)}}{r^2}\int_0^{r}t^2(e^{-\alpha S(t)}w^q(t)-\sigma)dtdr\notag\\
 &\ge p-\int_0^{r_1}\frac{e^{-\alpha \underline {S}}}{r^2}\int_0^{r}t^2(e^{-\alpha \underline {S}}p^q-\sigma)dtdr\notag\\
 &= p-\frac{r_1^2}{6}(e^{-2\alpha \underline {S}}p^q-e^{-\alpha \underline {S}}\sigma),
 \end{align}
if $p$ is sufficiently large, where $\underline S=\inf_{0\le
r<+\infty} S(r),$ and by (2.15) $\underline S>-\infty$. From
(\ref{t1}), we get
\begin{equation}
\frac{r_1^2}{2}(e^{-2\alpha \underline {S}}p^q-e^{-\alpha \underline {S}}\sigma)\ge p.
\end{equation}
This implies
$$r_1\ge c_1 p^{-(q-1)/2}, $$
for some positive constant $c_1$, if $p$ is sufficiently large.
Similar to the argument in (\ref{t1}), we can obtain,
\begin{equation}\label{wed}
 2p/3\le p-\frac{r_1^2}{6}(e^{-2\alpha \bar {S}}(2p/3)^q-e^{-\alpha \bar {S}}\sigma),
 \end{equation}
 where
$\bar S=\sup_{0\le r<+\infty} S(r),$ and $\bar S<\infty$. So from
(\ref{wed}),
$$r_1\le c_2 p^{-(q-1)/2}, $$
for some positive constant $c_2$, if $p$ is sufficiently large.
This proves (3.15). To prove (\ref{2.1x}), we have, in view of
(\ref{2.6}),
\begin{equation}\label{navier}
G(r_1)=\int_0^{r_1}
3r^2\left\{[\frac{1}{q+1}-\frac{1}{6}-\frac{\alpha}{3(q+1)}rS'(r)]e^{-\alpha
S}w^{q+1} -\frac{5}{6}\sigma w\right\}dr. \end{equation} Since
$1<q<5$, ($q+1<6$), we have
\begin{equation}\label{stokes}
\frac{1}{q+1}-\frac{1}{6}-\frac{\alpha}{3(q+1)}rS'(r)>c_3,
\qquad 0\le r\le r_1,
\end{equation}
if  $\sup_{0\le r<r_1}|S'(r)|$ is sufficiently small,
where $c_3$ is some positive constant independent of $p$. By virtue of
(\ref{t2}), (\ref{navier}) and (\ref{stokes}), we have,
\begin{align}\label{2.13}
&G(r_1)\ge r_1^3 \left(c_3e^{-\alpha \bar
S}(w(r_1))^{q+1}-\frac{5}{6}\sigma p
\right)\notag\\
&= r_1^3\left(c_3e^{-\alpha \bar S}(2p/3)^{q+1}-\frac{5}{6}\sigma
p \right).
\end{align}
Therefore,  if $p$ is sufficiently large,  (\ref{2.13}) implies
(\ref{2.1x}). This proves Lemma 3.1.
\end{proof}
By (\ref{f(w)}), we have
\begin{equation}\label{decrease}
f(r,\ w)>0, \ for~ w>[{e^{\alpha \bar S}\sigma}]^{1/q}.
\end{equation}
Therefore, in view of (\ref{2.11}),  we have $w'(r)<0$ for small
$r$ if $p$ is sufficiently large, so $w(r)$ decreases for small
$r$. Using (\ref{2.11}) and (\ref{decrease}), we can see  that
$w'(r)<0$ and thus $w(r)$ decreases as long as $w>[{e^{\alpha \bar
S}\sigma}]^{1/q}$. Because of this, we can define $T=:T(p)$ as the
point such that
\begin{equation}\label{T}
w(T)=[\frac{10(q+1)e^{\alpha \bar S}\sigma}{(5-q)}]^{1/q}=:A,\
w(r)>A,\ for\ 0\le r<T,
\end{equation}
since
\begin{equation}\label{a1}
w(T)=A
>[e^{\alpha \bar S}\sigma]^{1/q},
\end{equation}
because $1<q<5$. So by (\ref{f(w)}) and (\ref{T}), we have
$$f(r, w(r))>0,\ 0\le r\le T,$$
and thus
\begin{equation}\label{a2}
w'(r)<0, \qquad {\rm for~} 0<r\le T.
\end{equation}
\begin{rem} The existence of such a $T$  follows here by a similar argument as in \cite{SW}.
\end{rem}

The next lemma gives an upper bound for $T$, where the constant
$b$ is given by (\ref{b}).
\begin{lem}\label{boundT}
Assume that $1<q<5$ ($6/5<\gamma<2$) and
\begin{equation}\label{s'}
\alpha S'(r)e^{\alpha S}z'>-\frac{b}{2}z,
\end{equation}
for $0<r<\pi/\sqrt{b},$ where
\begin{equation}\label{z3}
z(r)=\frac{\sin(\sqrt{b}r)}{\sqrt{b}r}.
\end{equation}
  then
\begin{equation}\label{z5}
T\le \pi/\sqrt{b}.
\end{equation}
\end{lem}
\begin{proof}
It is easy to verify  $z(r)$ defined in (\ref{z3}) is  the
solution of the following
  initial value problem of second order linear equation
\begin{equation}\label{z}
z''+\frac{2}{r}z'+bz=0, \qquad z(0)=1,\ z'(0)=0,
\end{equation}
By (\ref{w'}) and (\ref{z}), we have
\begin{equation}\label{z4}
[r^2e^{\alpha S(r)}(z'w-w'z)]'=r^2[f(r,\ w)z -bze^{\alpha
S}w+\alpha S'(r)e^{\alpha S}z'w].
\end{equation}
We prove (\ref{z5}) by contradiction. Suppose $T>\pi/\sqrt{b}$, in
view of (\ref{f(w)}), (\ref{a1}) and (\ref{a2}), we have
\begin{align}\label{z6}
&\frac{f(r,\ w(r))}{w(r)}-be^{\alpha S(r)}\notag\\
&\ge e^{-\alpha \bar S}(w(T))^{q-1}-\frac{\sigma}{w(T)}-be^{\alpha
\bar S} =\frac{b}{2}.
\end{align}
Integrating (\ref{z4}) over the interval $[0,\  \pi/\sqrt{b}]$,
noticing the fact $z(\pi/\sqrt{b})=0$, we obtain
\begin{equation}\label{z7}
r^2e^{\alpha S(r)}z'w|_{r=\pi/\sqrt b}=\int_0^{\pi/\sqrt b}
r^2wz\left\{\frac{f(r,\ w)}{w} -be^{\alpha S}+\frac{\alpha
S'(r)e^{\alpha S}z'}{z}\right\}dr.
\end{equation}
In view of (\ref{a2}), we have $w(\frac{\pi}{\sqrt b})>w(T)>0$ if
$T> \frac{\pi}{\sqrt b}$. Moreover from (3.29)
$z'(\frac{\pi}{\sqrt b})<0$. Thus the left hand side of (\ref{z7})
is negative. On the other hand, by (\ref{s'}) and (\ref{z6}), we
can see that the right hand side of (\ref{z7}) is positive. This
is a contradiction, and thus (\ref{z5}) is proved.
\end{proof}
In order to prove Parts 2) and 3) in Theorem 2.1, we need a few
lemmas.
 First, in view of (\ref{f(w)}) and (\ref{F(w)}),
there exists a positive constant $B$  such that
\begin{equation}\label{A1}
|e^{-\alpha S(r)} F(r,\ w)|\le B, {\rm ~for~} 0\le w\le A,
\end{equation}
\begin{equation}\label{A2}
e^{-\alpha S(r)} |g(r,\ w)- \alpha S'(r) F(r,\ w)|\le B, {\rm
~for~} 0\le w\le A,
\end{equation}
where $A=w(T)$ is  defined in (\ref{T}),  $g(r,\ w)$  is  defined
in (\ref{g}). In the following, we denote
\begin{equation}\label{z12}
Q:=Q(T)=w'(T),
\end{equation}
where $T=T(p)$ is defined in (\ref{T}). By (\ref{a2}), we have
\begin{equation}\label{z13}
Q<0.
\end{equation}
\begin{lem}\label{lemay161}
If the entropy $S$ satisfies the hypothesis in Lemma 3.2, and
\begin{equation}\label{e1}
\frac{4B}{Q^2}(1+\frac{A}{|Q|})+\frac{12A}{|Q|T}(1+\frac{4B}{Q^2}+\frac{4AB}{|Q|^3})<\frac{1}{2},
\end{equation}
and
\begin{equation}\label{apriori}
0\le w(r)\le A=w(T), {\rm ~for~} T\le r\le T+L,
\end{equation}
for any $L$ satisfying
\begin{equation}\label{apriori1}
0\le L\le \min\{\frac{2A}{|Q|},\ 1\},
\end{equation}
then
\begin{equation}\label{e2}
w'(r)\le -\frac{|Q|}{\sqrt 2}, {\rm~for~} T\le r\le T+L,
\end{equation}
provided $\sup_{0\le r<1+\frac{\pi}{\sqrt b}} |S'(r)|$ is
sufficiently small.
\end{lem}
\begin{proof}
First, by (\ref{eq3.3}), we have
\begin{equation}\label{z11}
(\frac{w'^2}{2}+e^{-\alpha S}F(r,\ w))'=-\frac{2+\alpha
rS'(r)}{r}w'^2 +e^{-\alpha S(r)} [g(r,\ w)- \alpha S'(r) F(r,\
w)],
\end{equation}
where $g(r,\ w)$ is given by (\ref{g}). Notice that $T$ is bounded
by $\pi/\sqrt b$ (cf. (2.30)), where $b$ is given by (\ref{b}), so
\begin{equation}\label{z15}
 [T, T+1]\subset [0, \frac{\pi}{\sqrt b}+1].
\end{equation}
This, together with (\ref{apriori1}), implies
\begin{equation}\label{z16}
[T,\ T+L]\subset  [0, \frac{\pi}{\sqrt b}+1].
\end{equation}

Therefore, if $\sup_{0\le r<1+\frac{\pi}{\sqrt b}}|S'(r)|$ is
sufficiently small, then $|rS'(r)|$ is also small for $T\le r\le
T+L$. Thus
 \begin{equation}\label{z17}
0<2+\alpha rS'(r)<3,
\end{equation}
for $T\le r\le T+L$. It follows from (\ref{z11}) and (\ref{z17})
that,
\begin{align}\label{z18}
 &\frac{w'^2}{2}+e^{-\alpha S}F(r,\ w)\notag\\
&\le  \frac{Q^2}{2}+e^{-\alpha S(T)}F(T,\ w(T))+\int_T^r
[e^{-\alpha S(t)} (g(t,\ w)- \alpha S'(t) F(t,\ w)]dt,
\end{align}
for $T\le r\le T+L$. Using (\ref{A1}), (\ref{A2}) and (\ref{z18}),
we obtain
\begin{equation}\label{z22}
 \frac{w'^2}{2}
\le  \frac{Q^2}{2}+2B+BL\le \frac{Q^2}{2}+2B+\frac{2BA}{|Q|}
\end{equation}
for $T\le r\le T+L$.
 On the other hand, we have from (3.36) and (3.43)   that
\begin{equation}\label{z24}
 \frac{w'^2}{2}+e^{-\alpha S(r)} F(r, w)
\ge \frac{Q^2}{2}+e^{-\alpha S(T)} F(T,
w(T))-\int_T^r\frac{2+\alpha tS'(t)}{t}(w'(t)^2dt-BL,
\end{equation}
for $T\le r\le T+L$. By (3.35), (3.46) and (3.49), we obtain
\begin{align}
&\frac{w'^2}{2} \ge
\frac{Q^2}{2}-2B-BL-\int_T^{T+L}\frac{3}{t}(w'(t)^2dt \notag\\
&\ge
\frac{Q^2}{2}-2B(1+\frac{L}{2})-\int_T^{T+L}\frac{3}{T}(w'(t)^2dt,
\end{align}
for $T\le r\le T+L$. So, if $L$ satisfies (3.41), by virtue of
(3.48), we have
\begin{align}\label{z25}
& \frac{w'^2}{2} \ge
\frac{Q^2}{2}-2B(1+\frac{A}{|Q|})-\int_T^{T+\frac{2A}{|Q|}}\frac{3}{T}(w'(t)^2dt\ \ \ ({ ~by~ (3.41)})\notag\\
&\ge\frac{Q^2}{2}-2B(1+\frac{A}{|Q|})-\int_T^{T+\frac{2A}{|Q|}}\frac{3}{T}
\left(Q^2+4B+\frac{4BA}{|Q|}\right)dt\ \ ({~by~ (3.48)})\notag\\
&= \frac{Q^2}{2}-2B(1+\frac{A}{|Q|})-\frac{6A}{T|Q|}(
Q^2+4B+\frac{4BA}{|Q|}),
\end{align}
for $T\le r\le T+L$. Hence
\begin{equation}\label{z25}
 \frac{w'^2}{Q^2}
\ge  1-\frac{4B}{Q^2}(1+\frac{A}{|Q|})-\frac{12A}{T|Q|}(
1+\frac{4B}{Q^2}+\frac{4BA}{|Q|^3}),
\end{equation}
for $T\le r\le T+L$. Therefore, if (\ref{e1}) holds, then we have
\begin{equation}\label{z27}
 (w'(r))^2
\ge \frac{Q^2}{2}
\end{equation}
for $T\le r\le T+L$. This implies $w'(r)$ does not change sign for
$T\le r\le T+L$.  Since $Q=w'(T)<0$ (cf. (\ref{a2})),
\begin{equation}\label{z27}
 w'(r)
\le -\frac{|Q|}{\sqrt 2}
\end{equation}
for $T\le r\le T+L$.
\end{proof}
Next, we have the following lemma.
\begin{lem}\label{lemay162}
If $w(r)>0$ for $r\in [T,\ T+L]$ where  $L$ satisfies
(\ref{apriori1}), then we have
\begin{equation}\label{e3}
0<w(r)<A, \  for~ T<r\le T+L.
\end{equation}
\end{lem}
\begin{proof}
Since $w(T)=A$ and $w'(T)=Q<0$ as we showed before, then
$w'(r)<0$, and thus $w(r)<A$ for $r>T$, $(r-T)$  small. We prove
(\ref{e3}) by contradiction. If (\ref{e3}) were false, then there
exists $r_2\in (T, T+L]$ such that
\begin{equation}
 w(r)<A {~\rm for~} r\in (T, r_2),\   w(r_2)=A.
 \end{equation}
 Since $w(T)=w(r_2)=A$ and $r_2>T$, by Rolle's Theorem, we  have
 \begin{equation}
 w'(\tau)=0,
 \end{equation}
 for some $\tau\in (T, r_2)$. This contradicts (\ref{z27}). The proof of the lemma is complete.
 \end{proof}
 The following lemma is a generalization of a result in \cite{SW}.

 \begin{lem}\label{sw1} Assume the entropy $S$ satisfies the hypothesis  in Lemma \ref{boundT} .
 Let $T(p)=T$ be the point defined in (\ref{T}).
If
\begin{equation}\label{condition}
 w'(T)T\to -\infty,  {\rm~as~}
p\to +\infty, \end{equation}  then $R(p, \sigma)<+\infty$ if $p$
is sufficiently large and $\sup_{0\le r\le 1+\frac{\pi}{\sqrt b}
}|S'(r)|$ is sufficiently small.
\end{lem}
\begin{proof}
First,  in view of  (\ref{z5}), (\ref{condition}) implies
\begin{equation}\label{z14}
Q(T)=w'(T)\to -\infty,  {\rm~as~} p\to +\infty. \end{equation}
Hence, there exists $p_0>0$ such that (3.59) holds for every $p\ge
p_0$. Now, for any fixed $p\in [p_0, +\infty)$, we show that there
exists $r^{*}\in [T, \ T+\frac{2A}{|Q|}]$ such that
\begin{equation}\label{m1}
w(r^{*})\le 0,
\end{equation}
and this implies $R(p,\ \sigma)<+\infty.$ We prove (\ref{m1}) by
contradiction. Suppose
\begin{equation}\label{m4}
w(r)>0\ {~\rm for~} r \in [T, \ T+\frac{2A}{|Q|}].
\end{equation}
Then by Lemma \ref{lemay162}, we have
\begin{equation}\label{m2}
0<w(r)<A, {~\rm for~} r \in [T, \ T+\frac{2A}{|Q|}].
\end{equation}
Thus, we can apply Lemma \ref{lemay161} with $L=T+\frac{2A}{|Q|}$
to obtain
\begin{equation}\label{m3}
w'(r) \le -\frac{|Q|}{\sqrt 2}, {~\rm~for~} T\le r\le
T+\frac{2A}{|Q|}.
\end{equation}
Therefore, since $w(T)=A$, we have
\begin{equation}\label{m5}
w(T+\frac{2A}{|Q|})=A+\int_T^{T+\frac{2A}{|Q|}}w'(r)dr\le
A-\frac{|Q|}{\sqrt 2}\cdot\frac{2A}{|Q|}<0.
\end{equation}
This contradicts (\ref{m4}). The proof of the lemma is complete.
\end{proof}

For the case $1<q<3$ ($4/3<\gamma<2$), we have the following
result. This is the same as Part 2) in Theorem 2.1. {\bf
Proposition 1} \begin{it} Assume $1<q<3$ ($4/3<\gamma<2$) and
$\sigma>0$. If the entropy $S(r)$ satisfies condition (\ref{s'})
in Lemma \ref{boundT}, then
\begin{equation}
R(p,\ \sigma)<+\infty,
\end{equation}
for sufficiently large $p$, provided $\sup_{0\le
r<1+\frac{\pi}{\sqrt b}}|S'(r)|$  is sufficiently small.
\end{it}\\

\begin{proof}
To prove this proposition, it suffices to verify (\ref{condition})
in Lemma \ref{sw1}. This follows by the following argument. Set
$$w'(T)=Q.$$
By (\ref{2.11}), we have
\begin{equation}\label{z30}
-e^{\alpha S(T)}T^2Q=\int_0^Tr^2f(r,\ w(r))dr.
\end{equation}
We estimate $QT^2$ as follows. For $0\le r\le T$, by  (\ref{a2}), we have
\begin{equation}\label{z31}
w(r)\ge w(T).
\end{equation}
Therefore, by (\ref{f(w)}) and (\ref{T}), we obtain
\begin{align}\label{z32}
f(r,\ w(r))&=e^{-\alpha S(r)}w^q(r)-\sigma \notag\\
&\ge e^{-\alpha \bar S}w^q(r)-\sigma \notag\\
&\ge e^{-\alpha \bar S}w^q(T)-\sigma \notag\\
&\ge \frac{(5+11q)\sigma}{5-q}>0,
\end{align}
for $0\le r\le T$, since $1<q<3$. On the other hand, for $r_1$
defined in (3.14),  since $2p/3>A=w(T)$  if $p$ is sufficiently
large, then  (3.25) and (\ref{a2}) imply
\begin{equation}\label{z33}
T>r_1,
\end{equation}
for $p$ large.
  Hence, it
follows from (\ref{z30})-(\ref{z33}) that
\begin{equation}\label{z34}
-e^{\alpha S(T)}T^2Q\ge\int_0^{r_1}r^2f(r,\ w(r))dr.
\end{equation}
By (3.14), we have
\begin{equation}\label{z35}
w(r)\ge w(r_1)\ge 2p/3, \qquad {\rm for~} 0\le r\le r_1.
\end{equation}
Thus, by (\ref{f(w)}),
\begin{equation}\label{z36}
f(r, w(r))\ge e^{\alpha \bar S}(3p/2)^q-\sigma\ge const\cdot p^q,
\qquad {\rm for~} 0\le r\le r_1.
\end{equation}
if $p$ is sufficiently large. Therefore, we obtain, by (\ref{r1}),
(\ref{z34}) and (\ref{z36}), that
\begin{equation}\label{z37}
-e^{\alpha S(T)}T^2Q\ge const\cdot p^qr_1^3=\ge const\cdot
p^{\frac{3-q}{2}}.
\end{equation}
This implies $T^2Q\to -\infty$ as $p\to +\infty$ if $q<3$.
Condition (\ref{condition}) is thus verified in view of (\ref{z5})
in Lemma \ref{boundT}.
\end{proof}

For the case $1<q<5$ ($6/5<\gamma<2$), we have the following
result. This is the same as Part 3) in Theorem 2.1. \\

\noindent\begin{it} {\bf Proposition 2.} Assume $1<q<5$
($6/5<\gamma<2$) and $\sigma>0$. If the entropy $S(r)$ satisfies
the hypothesis in Proposition 1 and
\begin{equation}\label{z38}
S'(r)\le 0, {\rm~ for~} 0\le r\le \frac{\pi}{\sqrt b},
\end{equation}
where $b$ is given by (\ref{b}) , then
\begin{equation}
R(p,\ \sigma)<+\infty,
\end{equation}
for sufficiently large $p$, provided $\sup_{0\le
r\le 1+\frac{\pi}{\sqrt b}}|S'(r)|$  is sufficiently small.
\end{it}\\
\begin{proof}
First, in view of (\ref{t2}) and (\ref{T}), we have
\begin{equation}\label{191}
w(r_1)>w(T),
\end{equation}
if $p$ is sufficiently large. This, together with (\ref{a2}),
implies
\begin{equation}\label{192}
r_1<T.
\end{equation}
Once again, by (\ref{T}),  we obtain
\begin{equation}\label{193}
w(r)\ge w(T),
\end{equation}
for $r_1\le r\le T$.  It follows from  (\ref{2.5})  that
\begin{equation}\label{2.17}
G(T)=G(r_1)+\int_{r_1}^T 3r^2\left(F(r,\ w)-\frac{1}{6}f(r,\
w)w+\frac{1}{3}rg(r,\ w)\right)dr.
\end{equation}
By  (\ref{f(w)}), (\ref{F(w)}), (\ref{g}), we have
\begin{align}\label{194}
&F(r,\ w)-\frac{1}{6}f(r,\ w)w+\frac{1}{3}rg(r,\ w)\notag\\
&=\left\{\left[\frac{1}{q+1}-\frac{1}{6}-\frac{\alpha
rS'(r)}{3(q+1)}\right] e^{-\alpha
S}(w(r))^q-\frac{5}{6}\sigma\right\}w(r),
\end{align}
for $r_1\le r\le T$. If $\sup_{0\le r\le 1+\frac{\pi}{\sqrt
b}}|S'(r)|$ is small, in view of (\ref{z5}), we have
\begin{equation}\label{195}
\frac{1}{q+1}-\frac{1}{6}-\frac{\alpha rS'(r)}{3(q+1)}>
\frac{1}{2}\left(\frac{1}{q+1}-\frac{1}{6}\right)>0,
\end{equation}
for $r\le T$ and $1<q<5$. Moreover, by (\ref{T}), (\ref{2.17}),
(\ref{193}) and (\ref{195}), we obtain
\begin{align}\label{196}
&F(r,\ w)-\frac{1}{6}f(r,\ w)w+\frac{1}{3}rg(r,\ w)\notag\\
&=\left\{\frac{1}{2}\left[\frac{1}{q+1}-\frac{1}{6}\right]
e^{-\alpha S}(w(T))^q-\frac{5}{6}\sigma\right\}w(r)\ge 0,
\end{align}
for $r_1\le r\le T$. This, together with (\ref{2.1x}) and
(\ref{2.17}),   implies
\begin{equation}\label{197}
G(T)\ge G(r_1)\ge c_3 p^{(5-q)/2}.
\end{equation}
By (2.5) and (\ref{197}), we have
\begin{align}\label{198}
&\frac{e^{\alpha S(T)}T}{2}([TQ]^2+(T)^3F(T, w(T))
+\frac{TAe^{\alpha S(T)}}{2}[TQ]+
\frac{\alpha}{2}\int_0^Tr^3e^{\alpha S(r)}S'(r)(w'(r))^2dr\notag\\
&\ge c_3p^{(5-q)/2},
\end{align}
where $Q=w'(T)$. So, if $S'(r)\le 0$ for $0\le r\le
\frac{\pi}{\sqrt b},$  we have, in view of (\ref{z5}),
\begin{align}\label{199}
&\frac{e^{\alpha S(T)}T}{2}[TQ]^2+(T)^3F(T, w(T))
+\frac{TAe^{\alpha S(T)}}{2}[TQ]\notag\\
&\ge c_3p^{(5-q)/2},
\end{align}
Since $T\le \frac{\pi}{\sqrt b}$ and $1<q<5$, we have
\begin{equation}
TQ\to -\infty, {~\rm as~} p\to +\infty.
\end{equation}
By virtue of Lemma \ref{sw1},  Proposition 2  is proved.
\end{proof}
Now we prove Part 4) of Theorem 2.1. For the solutions $w(r,\ p,\
\sigma)=:w(r)$ in 2) and 3) of Theorem 2.1, let $\rho=
\left[\frac{\gamma-1}{\gamma}w\right]^{1/(\gamma-1)}$;  $\rho$
is the density function (cf. (2.3)). Set
$$M(r)=\int_0^r4\pi \tau^2\rho(\tau)d\tau,$$
  the  mass in the ball $B_r(0)$.
  We calculate each term in (2.5) as follows, by virtue of the fact that $w$ and $S$ are spherically symmetric and
  $\Omega$ is a constant. First,  for
  $r=\sqrt{x_1^2+x_2^2+x_3^2}$, we have
\begin{equation}\label{june51}
\nabla w=w_r(x_1/r,\  x_2/r, \ x_3/r),\ \ \nabla S=S_r(x_1/r,\
x_2/r, \ x_3/r),
\end{equation}
and thus
\begin{align}\label{june52}
div(e^{\alpha S} \nabla w)&=\nabla (e^{\alpha S})\cdot \nabla
w+e^{\alpha S}\Delta w\notag\\
&=\alpha e^{\alpha S}w_rS_r+e^{\alpha
S}(w_{rr}+\frac{2}{r}w_r)\notag\\
&=e^{\alpha S}w_{rr}+e^{\alpha S}(\frac{2}{r}+\alpha S'(r))w_{r}.
\end{align}
By (2.3) and (2.6), we have
\begin{equation}\label{june53}
Ke^{-\alpha S} w^q=4\pi G\rho.
\end{equation}
Substituting (\ref{june52}) and (\ref{june53}) into (2.5), and
noticing $\Omega$ is a constant, we obtain
\begin{equation}
e^{\alpha S}w_{rr}+e^{\alpha S}(\frac{2}{r}+\alpha
S'(r))w_{r}+4\pi G\rho-2\Omega^2=0.
\end{equation}
Thus
\begin{equation}
  (r^2e^{\alpha S} w'(r))'+r^2(4\pi G \rho -2\Omega^2)=0,
  \end{equation}
  where $G$ is the Newtonian gravitational constant.  Therefore
  \begin{equation}\label{7082}
  r^2e^{\alpha S} w'(r)=\frac{2r^3\Omega^2}{3}-GM(r),
  \end{equation}
  for $r\le R(p,\ \sigma)$.  By  the proofs of Part 2) and 3) of Theorem 2.1, we have $w'(r)\le 0$ for $r\le  R(p,\ \sigma)$.
  Thus  $M(r)\ge \frac{2r^3\Omega^2}{3G}$,  for $r\le  R(p,\ \sigma)$.  The estimate (\ref{adensity}) follows immediately from
  (\ref{mass}).

  \section{Proofs of Theorems \ref{thm4} and \ref{thmx} }
In this section,  we consider the isentropic case $S=constant$.
Theorem 2.1 can be applied in this case because the conditions
imposed on the entropy $S$ in Theorem 2.1  are automatically
satisfied when $S$ is constant.  Without loss of generality,
throughout this section, we assume
\begin{equation}
S=0 \end{equation}
 for convenience.  We
prove Theorems \ref{thm4} and \ref{thmx} as follows. First, we
define
\begin{equation}\label{f(w1)}
f(w)=w^q-\sigma,\  w\ge 0,
\end{equation}
and
\begin{equation}\label{F(w1)}
F(w)=\frac{w^{q+1}}{q+1}-\sigma w,\  w\ge 0
\end{equation}

We still use $w(r, p, \sigma)$  to denote the solution of the
problem
\begin{equation}\label{2.3}
w''(r)+\frac{2}{r}w'(r)+f(w)=0,
\end{equation}
and
\begin{equation}\label{2.4}
w'(0)=0, w(0)=p>0.
\end{equation}
  Let $R(p,\ \sigma)$ be defined as in (2.13); $R(p,\ \sigma)$
is the first point at which $w$ is 0.  Then we have the following
proposition.

 \noindent {\it {\bf Proposition 3.} If $\sigma> 0$ and
$1<q<5$, ($6/5<\gamma<2$),  then
\begin{equation}
R(p,\sigma)=+\infty,  {~\rm if~} 0\le p< \underline{p} ,
\end{equation}
where
\begin{equation}\label{pp}
\underline{p}=(\frac{5\sigma(1+q)}{5-q})^{1/q}=(\frac{10\gamma\Omega^2}{5\gamma-6})^{\gamma-1}.
\end{equation}}

\vspace{0.5cm}

\begin{proof}
It is easy to verify that $(\frac{5\sigma(1+q)}{5-q})^{1/q}$ is
the unique positive zero of $F(w)-\frac{f(w)w}{6}$. If $R(p,
\sigma)<+\infty$, we let $R(p, \ \sigma)=R>0$. Then $G(R)\ge 0$,
where the function $G$ is defined as in (2.5) with $S=0$. On the other hand
$0\le w(r)\le p$ for $0\le r\le R$. If
$p<\underline{p}=(\frac{5\sigma(1+q)}{5-q})^{1/q}$, then
$(F(w)-\frac{f(w)w}{6})(r)< 0$ for $0\le r\le R$. Hence,
(\ref{2.6}) implies $G(R)< 0$. This is a contradiction.
\end{proof}

\noindent{\it Proof of Part 1) of Theorem 2.2}\\
For fixed $\sigma>0$, let
\begin{equation}\label{ww}
p_0=\inf\{p:\ R(p, \sigma)<+\infty\}.\end{equation} By Part 3) of
Theorem 2.1 and
 Proposition 3,  we know that
$$\underline{p}\le p_0<+\infty,$$
where $\underline{p}$ is given by (\ref{pp}). If we can show
\begin{equation}\label{p0}
R(p_0, \sigma)<\infty,
\end{equation}
then Part 1) in Theorem 2.2  will be proved. Now (\ref{p0}) can be
shown by the following argument. Write (\ref{2.3}) as a first
order system
\begin{equation}\label{system}
w'=v,\qquad v'=-\frac{2}{r}v+\sigma-w^q,
\end{equation}
with initial condition
\begin{equation}\label{ic}
 w(0)=p_0>0,\  v(0)=0.
\end{equation}
Define the Hamiltonian $H(w,\ v)$ by
\begin{equation}\label{hom}
H(w, v)=\frac{v^2}{2}+\frac{w^{q+1}}{q+1}-\sigma w.
\end{equation}
Then $H'=-\frac{2}{r}v^2$, so that $H$ decreases on orbits of
 (\ref{system}). This implies that the solution of (\ref{system})-
(\ref{ic}) must have $v$ bounded from below in the region
$S=\{w\ge 0,\ v\le 0\}$. This solution also cannot exit $S$ via
$w=0$ at some $v<0$ for some $R>0$; otherwise, since
$v'(R)=-\frac{2}{R}v(R)+\sigma>0,$ the solution crosses the line
$w=0$ transversally, so by continuity, there would be a
neighborhood $N$ of $p_0$ on the $w$-axis, such that for $p\in N,$
the orbit of (\ref{system}) satisfying $w(0)=p$, $v(0)=0$, would
also exit $S$ at a point near $w=0,$  $v(R)<0$. This would
contradict the definition of $p_0$. Similarly the $p_0$ orbit
cannot exit $S$ via some point $(w,\ 0),$  with $0<w<p_0.$ Thus
the $p_0$ orbit exits $S$ via $w=0,$  $v=0,$  so (2.31) holds,
and this proves Part 1) of Theorem 1.2.\\

\noindent \begin{rem} It is not  hard to show that every solution
of (\ref{system}) satisfying $w(0)=p,$ $v(0)=0,$ tends to the rest
point $(w=\sigma^{1/q},\ v=0)$ as $r\to \infty.$\end{rem}

\vspace{0.5cm}

{\it Proof  of Part 2) of Theorem \ref{thm4}.}\\

First, we show that the radius of the star increases with the increasing angular velocity. \\
 We use $w_i(r)$
($i=1,\ 2$) to denote the solution to the following initial value
problem
\begin{equation}\label{eq4.13}
\begin{cases}
&w_i''(r)+\frac{2}{r}w_i'(r)+w_i^q-\sigma_i=0,\qquad r>0, \\
& w_i(0)=p,\ w_i'(0)=0. \end{cases}
\end{equation}
We assume
\begin{equation}\label{eq2}
\sigma_1>\sigma_2> 0,
\end{equation}
and we want to show that
\begin{equation}\label{eq3}
R(p,\ \sigma_1)\ge R(p,\ \sigma_2).
\end{equation}
(Note that each of these are finite, by what we have already shown in Part 2) of the theorem).

From (\ref{eq4.13}), we have
 \begin{equation}\label{wi}
w_i'(r)=\frac{1}{r^2}\int_0^r
s^2\left(\sigma_i-w_i^q(s)\right)ds,\qquad i=1,\ 2,
\end{equation}
It is easy to verify,  using L'Hospital's rule,  that
\begin{equation}\label{eq4}
\lim_{r\to 0+}\frac{2}{r}w_i'(r)=\frac{2}{3}(\sigma_i-p^q).
\end{equation}
It follows from (\ref{eq4.13}) and (\ref{eq4}) that
\begin{equation}\label{eq5}
\lim_{r\to 0+}w_i{''}(r)=-(p^q-\sigma_i)/3.
\end{equation}
Since $\sigma_1>\sigma_2$, (\ref{eq5}) implies
\begin{equation}\label{eq6}
w{''}_1(0+)>w{''}_2(0+).
\end{equation}
This, together with the fact $w_1(0)=w_2(0)=p$ and
$w_1'(0)=w_2'(0)=0$, leads to
\begin{equation}\label{eqz}
w_1(r)>w_2(r), \qquad {\rm for~} {~\rm small~} r>0.
\end{equation}
We shall show (\ref{eq3}) by contradiction. If (\ref{eq3})
were false, then there exists an $r_0$  $0<r_0<R(p, \sigma_1)<R(p,
\sigma_2)$ such that
\begin{equation}\label{eq7}
w_1(r)>w_2(r),\ {\rm~for~} 0<r<r_0, {\rm~and~}  w_1(r_0)=w_2(r_0).
\end{equation}
Let
$$y(r)=w_1(r)-w_2(r).$$
We then have from (4.13),  that
\begin{equation}\label{eq8}
y{''}(r)+\frac{2}{r}y'(r)+yB(r)-(\sigma_1-\sigma_2)=0.
\end{equation}
Here \begin{equation}\label{B(r)}
 B(r)=q\int_0^1\left(\lambda
w_1+(1-\lambda)w_2\right)^{q-1}(r)d\lambda. \end{equation}
From (\ref{wi}) and (\ref{eq7}),  we have
\begin{equation}\label{eq9}
y(0)=y'(0)=0,\ y(r)>0 \ {\rm~for,~}\  0<r<r_0\ {\rm and~} y(r_0)=0.
\end{equation}
Multiplying (\ref{eq8}) by $y'$ and integrating the resulting
equation over the interval $[0,\ r_0]$, we get, since $y'(0)=0$,
\begin{align}\label{eq10}
&\frac{(y'(r_0))^2}{2}+\int_0^{r_0}\frac{2}{r}(y'(r))^2dr\notag\\
&+\int_0^{r_0}
B(r)yy'(r)dr-\int_0^{r_0}(\sigma_1-\sigma_2)y'(r)dr=0.
\end{align}
 With the help of (\ref{eq9}) and
integration by parts, we get
\begin{equation}\label{eq11}
\int_0^{r_0}B(r)yy'(r)dr=\int_0^{r_0}B(r)(\frac{y^2}{2})'dr=-\frac{1}{2}\int_0^{r_0}B'(r)y^2(r)dr
\end{equation}
and
\begin{equation}\label{eq12}
\int_0^{r_0}(\sigma_1-\sigma_2)y'(r)dr=0.
\end{equation}
substituting (\ref{eq11}) and (\ref{eq12}) into (\ref{eq10}), we have
\begin{equation}\label{eq15}
\frac{(y'(r_0))^2}{2}+\int_0^{r_0}\frac{2}{r}(y'(r))^2dr
+\frac{1}{2}\int_0^{r_0} (-B'(r))y^2(r)dr=0.
\end{equation}
By the definition of $B(r)$ (see (\ref{B(r)})), we have
\begin{equation}\label{eq16}
B'(r)=q(q-1)\int_0^1(\lambda w_1+(1-\lambda) w_2)^{q-2}(\lambda
w_1'(r)+(1-\lambda)w_2'(r))d\lambda.
\end{equation}
Since $w_i'(r)<0$ for $0<r<R(p, \sigma_i)$ ($i=1,\ 2$)
(\cite{Ni}), we thus have
\begin{equation}\label{eq17}
B'(r)<0,\  {\rm for~} 0<r\le r_0,
\end{equation}
when $\gamma< 2$, i.e., $q> 1$. Hence , each term in
(\ref{eq15}) must be zero. This contradicts (\ref{eq9}), and proves (\ref{eq3}).\\

Now we show that the radius of the star decreases with the increasing
central density, i.e.,
\begin{equation}\label{radiuss}
R(p_1, \sigma)\ge R(p_2, \sigma),
\end{equation}
if $p_0\le p_1<p_2$,  and $\sigma>0$. For this purpose, let $w(r,
p ,\sigma)$ be the solution of the following initial value problem

\begin{equation}\begin{cases}
& w''(r)+\frac{2}{r}w'(r)+w^q-\sigma=0, \\
& w(0)=p,\  w'(0)=0,
\end{cases}
\end{equation}

\noindent for $p\ge p_0$. We use the following rescaling,
\begin{equation}\label{scalling}
\lambda=rp^{(q-1)/2},\  \theta(\lambda)=w/p\ ;
\end{equation}
then $\theta(\lambda)$ is the solution of the following initial value problem:

\begin{equation}\label{x4}
\begin{cases}
&\theta_{\lambda\lambda}+\frac{2}{\lambda}\theta_{\lambda}+\theta^q-\frac{\sigma}{p^q}=0,\
\lambda>0, \notag\\
 &\theta (0)=1,\  \theta'(0)=0.
 \end{cases}
 \end{equation}

\noindent The first zero of $\theta$ depends only on the parameter
$\frac{\sigma}{p^q}$. We use $\lambda(\frac{\sigma}{p^q})$ to
denote this first zero. Then by (\ref{scalling}), we have
\begin{equation}\label{nnnn}
R(p, \sigma)=p^{(1-q)/2}\lambda(\frac{\sigma}{p^q}),
\end{equation}
for $p\ge p_0$.
Similar to the argument in the proof of (\ref{eq3}), we can show
that $\lambda(\frac{\sigma}{p^q})$ increases with the parameter
$\frac{\sigma}{p^q}$. Thus, for the fixed $\sigma$, it decreases
with $p$ for $p\ge p_0$. This, together with (\ref{nnnn}) and the
fact $q>1$, implies (\ref{radiuss}). This completes the proof of
Part 2) of Theorem 1.2.\\

\begin{rem}
The above scaling argument also works for non-rotating star, i.e.,
the case when $\sigma=0$. For the non-rotating star, the radius of
the star, $R(p, 0)$, is always finite for $p>0$, if $6/5<\gamma<2$
(see \cite{ch}). For the non-rotating star, (\ref{nnnn}) becomes
\begin{equation}\label{non}
R(p, 0)=\lambda_0p^{(1-q)/2},\qquad q=\frac{1}{\gamma-1},
\end{equation}
where $\lambda_0$ is the first zero of the function $\theta(\lambda)$, which is the solution
of the following initial value problem,
\begin{equation}\label{x5}
\begin{cases}
&\theta_{\lambda\lambda}+\frac{2}{\lambda}\theta_{\lambda}+\theta^q=0,\
\lambda>0, \notag\\
 &\theta (0)=1,\  \theta'(0)=0.
 \end{cases}
 \end{equation}
From (\ref{non}) we see that the radius of a non-rotating star is
proportional to $p^{(1-q)/2}$.
\end{rem}

We now turn to the proof of Theorem 2.3. In  (\ref{nnnn}), since
$\lambda(\frac{\sigma}{p^q})$ increases with the parameter
$\frac{\sigma}{p^q}$, as we have already shown, we have
\begin{equation}\label{nnnn1}
\lambda(\frac{\sigma}{p^q})\le \lambda(\frac{\sigma}{p_0^q}),
\end{equation}
for $p\ge p_0$. We apply (\ref{nnnn}) to the case of $p=p_0$ and obtain,
\begin{equation}\label{nnnn2}
R(p_0, \sigma)=p_0^{(1-q)/2}\lambda(\frac{\sigma}{p_0^q}).
\end{equation}
Now $R(p_0, \sigma)<+\infty$, from Part 2) of Theorem 2.1, and
$p_0$ is a positive constant determined completely by $\gamma$ and
$\sigma$. Thus $\lambda(\frac{\sigma}{p_0^q})$ is a positive
constant determined also only by $\gamma$ and $\sigma$. We set
this positive constant as $C$ in (\ref{nnnn1}), then (\ref{sig4})
follows. Since $\gamma<2$, (\ref{sig4}) implies (\ref{sig5}). The
proof of Theorem 2.2 is
complete.\\

The following lemma will be useful in proving Theorem
\ref{thm5}.
\begin{lem}\label{lem2.10}
 For any $\sigma> 0$, there exists $R_0>0$  such that (\ref{5}) with the data $w'(0)=0$ and $w(R)=0$  has a positive
solution in the ball $B_{R}(0)$ for all $R\le R_0$, if $1<q<5$, ($6/5<\gamma<2$).
\end{lem}
\begin{proof}
Fix $\sigma> 0$, and write $R(p, \ \sigma)=R(p)$. By Theorem
\ref{thm4} Part (1), there exists $p_0>0$ such that $R(p)<+\infty$
for any $p\ge p_0$. Let $R_0=R(p_0)$. For any $0<R<R_0$, it
follows from (\ref{sig5}) that, there exits $p_1>p_0$ such that
$R(p_1)<R$. Now  $R(p_0)>R>R(p_1)$, and $R(p)$ is a continuous
function of $p$ (actually, it is differentiable; cf. \cite{SW}).
Thus, by the intermediate value theorem, there exists $p\in (p_0,
p_1)$ such that $R(p)=R.$
  \end{proof}

 \section{Proofs of Theorems \ref{thm5}, 2.5 and 2.6}
 In this section, we consider the case for the general domains and variable
angular velocity. In Theorems 2.4 and 2.5, we assume the entropy function $S(x)$ is
constant.  We set
$S=0$ for convenience. Substitute this in (\ref{3}),
we get
\begin{equation}\label{3'}
\Delta w+w^q-2\Omega(\eta)( \Omega(\eta)+\eta \Omega'(\eta))=0,
\end{equation}
We look for the solution of (\ref{3'}) satisfying the following
condition
\begin{equation}\label{n2'}
w(x)>0, x\in D,\  w(x)=0, \  x\in \partial D.
\end{equation}
 In order to prove Theorem \ref{thm5}, we need a
comparison lemma, which can be found in
\cite{Evans} or \cite{smoller} . Before we state this lemma, we give the following
definitions of weak sub- and supersolutions of the problem (\ref{3'})
and (\ref{n2'}). For notational convenience, define
the function $f(x, w)$ by
$$f(x,
w)=w^q-2\Omega(\eta(x))[\Omega(\eta(x))+\eta(x)\Omega'(\eta(x)].$$
 \\
{\it Definition. (a)  $\bar w\in H^1(D)$  is called a weak
supersolution of problem (\ref{3'}) and (\ref{n2'}) if
\begin{equation}\label{3.1}
\int_{D}\nabla {\bar w}\cdot \nabla {v}dx\ge \int_{D}f(x, \bar
w)vdx
\end{equation}
for each $v\in C^1_0(D)$, $v\ge 0$ a.e., where the subscript zero denotes
$v=0$ on $\partial D$. Similarly\\
(b) $\underline w\in H^1(D)$  is called a weak subsolution
of problem (\ref{3'}) and (\ref{n2'}) if
\begin{equation}\label{3.2}
\int_{D}\nabla {\underline w}\cdot \nabla {v}dx\le \int_{D}f(x,
\underline w)vdx
\end{equation}
for each $v\in C^1_0(D)$, $v\ge 0$ }\\

The following lemma is well-known; cf \cite{Evans} or \cite{smoller}.

\begin{lem}\label{lem3.1}
Assume there exists a weak supersolution $\bar w$ and a weak
subsolution $\underline w$ of (\ref{3'}) and (\ref{n2'}) satisfying
\begin{equation}\label{3.3}
\underline w\le 0,\ \bar w\ge 0 ~ on~ \partial D ~in~
the~trace~sense,\ ~and~  \underline w\le \bar w\ a.e.~in~ D.
\end{equation}
Then there exists a weak solution $w\in H^1_0(D)$  of (\ref{3'}) and (\ref{n2'}), such
that
\begin{equation}\label{3.4}
\underline w\le w\le \bar w\qquad a.e.\ in\ D.
\end{equation}
\end{lem}
\begin{rem} By the smoothness assumptions of $\partial D$ and $\Omega(\eta)$, the usual
regularity arguments (see \cite{GT})  show that a weak positive solution of
(\ref{3'}) and (\ref{n2'}) must be a classical solution.\end{rem}

Let $D$ be a bounded domain with smooth boundary, and assume $\Omega(\eta)\in C^{1+\delta}(D)\cap C(\bar D)$,
for some $\delta$, $0<\delta<1$. Define $\beta$ by
 \begin{equation}\label{3.4'}
\beta=:\max_{{x}\in \bar D}|2\Omega (\eta)(\Omega
(\eta)+\eta\Omega'(\eta)|,
\end{equation}
where $\eta=\eta({ x})$ is as defined
in Section 1. Let $B_{R_1}(0)$ be the biggest ball contained in $D$.
Consider the positive solution of the equation
\begin{equation}\label{7}
u''(r)+\frac{2}{r}u'(r)+u^q-\beta=0,
\end{equation}
with data
\begin{equation}\label{8}
u'(0)=0,\  u(r)>0 {\rm  ~for~}  0<r<R_1,\   u(R_1)=0.
\end{equation}
By Lemma \ref{lem2.10}, if  $6/5<\gamma<2$, then there exists
a positive constant $R_2$ depending only on $\beta$ and $q$ such that
(\ref{7}) and (\ref{8}) has a solution if  $0<R_1\le R_2$.
 We require
\begin{equation}\label{R22}
 D\subset B_{R_2}(0).
\end{equation}
 This implies that (\ref{7}) and (\ref{8}) has a positive
solution. We  denote this solution by $u(x)=u(|x|)$
($x\in B_{R_1}(0)$). A subsolution to (\ref{3'}) and (\ref{n2'}) can now
be constructed.  First define $\underline{w}(x)$ by
\begin{equation}\label{3.7}
\underline{w}(x)=\left\{\begin{array}{ll} u(|x|) & {\rm for~}  x\in
B_{R_1}(0) \\
0 & {\rm for~}  x\in D-B_{R_1}(0).
\end{array}
\right.
\end{equation}

\begin{lem}\label{lem3.3}
If Condition A in Section 1 holds, then $\underline{w}(x)$
defined by (\ref{3.7}) is a  weak subsolution to (\ref{3'}) and
(\ref{n2'}).
\end{lem}
\begin{proof}
First, by the definition of $\underline w$ , we
have
\begin{align}\label{3.8}
&\int_{D}\left(\nabla {\underline w}\cdot \nabla {v}-f(x,
\underline w)v\right)dx\notag\\
&=\int_{B_{R_1}(0)}(\nabla{u}\cdot\nabla v- f(x,
u)v)dx\notag\\
&+\int_{D-B_{R_1}(0)}2\Omega(\eta(x))[\Omega(\eta(x))+\eta(x)\Omega'(\eta(x)]v(x)dx
\end{align}
for each $v\in C^1_0(D)$, $v\ge 0$.   By Condition A in
Section 1, we have \begin{equation}\label{3.9}
\Omega(\eta(x))[\Omega(\eta(x))+\eta(x)\Omega'(\eta(x)]\le 0,
\end{equation}
for $x\in D- B_{R_1}(0)$. On the other hand,  since $u$ satisfies
(\ref{7}) in the ball $B_{R_1}(0)$,  we have by the divergence
theorem,
\begin{align}\label{3.10}
&\int_{B_{R_1}(0)}\nabla{u}\cdot\nabla vdx\notag\\
&=-\int_{B_{R_1}(0)}\Delta u vdx
+\int_{\partial B_{R_1}(0)}\frac{\partial u}{\partial \nu} vdS\notag\\
&\le \int_{B_{R_1}(0)}f(x, u) vdx +\int_{\partial
B_{R_1}(0)}\frac{\partial u}{\partial \nu} vdS,
\end{align}
where $\nu$ is the unit outer normal vector. Since $u'(R)\le 0$ ,
we have $\frac{\partial u}{\partial \nu}\le 0$ on $\partial
B_{R_1}(0)$, and since $ u(x)=\underline {w}(x)$ for $x\in  B_{R_1}(0)$, (\ref{3.9}) implies

\begin{equation}\label{3.11}
\int_{B_{R_1}(0)}\nabla{u}\cdot\nabla vdx \le
\int_{B_{R_1}(0)}f(x, \underline w)\cdot vdx,
\end{equation}
so that \begin{equation}
\int_{D}\left(\nabla {\underline w}\cdot \nabla {v}-f(x,
\underline w)v\right)dx\le 0,
\end{equation}
and this proves Lemma \ref{lem3.3} .
\end{proof}

We next construct a supersolution to (\ref{3'}) and (\ref{n2'}). Let $R_2$ be the positive number as in (\ref{R22}) such that
$D\subset B_{R_2}(0)$ and $\bar p$ be any positive number such that
\begin{equation}\label{barp}
\bar p>\max_{x\in \bar D}\underline{w}(x),
\end{equation}
where $\underline{w}$ is the subsolution  which we have already constructed.
 We consider the following boundary value problem
\begin{equation}\label{3.13}
\Delta \bar {w}+\bar w^q+\beta=0, {~\rm for~} x\in B_{R_2}(0),
\end{equation}
and
\begin{equation}\label{boundary}
\bar w|_{\partial B_{R_2}(0)}=\bar p.
\end{equation}
  By a
result in \cite{Lions}, there exists a unique positive solution
$\tilde w(x)$ to problem (\ref{3.13}) and (\ref{boundary}). Furthermore, by the maximum principle , we have
\begin{equation}\label{max}
\bar w(x)\ge \bar p, \qquad {\rm for~} x\in B_{R_2}(0).
\end{equation}
Therefore, by the choice of $\bar p$ (see (\ref{barp}), we have
\begin{equation}\label{luohong}
\bar w(x)\ge \underline{w}(x),\qquad {\rm for~} x\in \bar D.
\end{equation}
It is easy to verify that $\bar w$ satisfies (\ref{3.1}), i.e., $\bar w(x)$ is a supersolution of (\ref{3'}) and (\ref{n2'}).
 Then (\ref{3.3})  follows from (\ref{luohong}) and the properties of $\underline{w}$ and $\bar w$.
 Hence, by Lemma 4.1, problem (\ref{3'}) and (\ref{n2'}) has a solution $w$ satisfying $\underline{w}(x)\le w(x)\le \bar
 w(x)$ for $x\in D$. Inequality (\ref{10}) can be proved in the same way as (\ref{low}).
 Now we prove $w(x)>0$ for $x\in D$. Since $w(x)\ge \underline{w}(x)$ for $x\in D$,
 and $\underline{w}(x)>0$ if $x\in B_{R_1}(0)$, it suffices to show
 $w(x)>0$ as $x\in D-B_{R_1}(0)$. This can be proved by the strong maximum principle
 (\cite{GT}), using Condition A in Section 1. This completes the proof of
  Theorem \ref{thm5}.

  We now prove Theorem \ref{new}. For this, we recall that the solution $w(x)$ of   (\ref{3'}) and
  (\ref{n2'})  is  constructed by the following iteration (cf. \cite{Evans} or \cite{smoller}).
  Set $w_0(x)=\underline{w}(x)$, and then define $w_k$ $(k=0,\ 1,\ 2,\ \, \cdots)$
  inductively to be the unique weak solution of the linear boundary-value problem
  \begin{equation}\label{new1}
  \begin{cases}
  &-\Delta w_{k+1}+Cw_{k+1}=f(w_{k})+Cw_k, \qquad {\rm in~} D\notag\\
  &w_{k+1}=0,\qquad {\rm on~} \partial D,
  \end{cases}
  \end{equation}
  where $C=\max_{x\in \bar D}|f'(\bar w)|(x)$.
  Then it can be shown (cf. \cite{smoller})
  $$\underline{w}=w_0\le w_1\le\cdots\le w_k\le \cdots\le \bar w, \ a.e.\ in\ D.$$
  Set $w(x)=\lim_{k\to\infty} w_k(x)$; then $w(x)$ is the desired solution.
  Since $\underline{w}$ and $\bar w$ are axi-symmetric,  if $D$ is an axi-symmetric domain,
  it is not hard to show each $w_k$ ($k$=0, 1, 2, ....) is axi-symmetric; i.e.  depends only on
  $\eta$ and $z$.   Indeed, if
$\eta=\sqrt{x_1^2+x_2^2}$, and $z=x_3$, then the Laplacian transforms to
$$\partial_{\eta\eta}+\frac{1}{\eta}\partial_{\eta}+\partial_{zz}.$$ So $w_1$
satisfies the equation
\begin{equation}\label{sss}
-(\partial_{\eta\eta}w_1+\frac{1}{\eta}\partial_{\eta}w_1+\partial_{zz}w_1)+Cw_1=f(\underline{w})
+C\underline{w},
\end{equation}
together with the boundary condition $w_1=0$ on $\partial D$.
Since the right hand side of (\ref{sss}) depends only on $\eta$
and $z$, and since this Dirichlet problem, being linear has a
unique solution, it follows that $w_1$ is axi-symmetric.
Similarly, each $w_k$ is axi-symmetric. Thus $w(x)$ is
axi-symmetric. This completes the proof of Theorem \ref{new}.\\

We finally prove Theorem \ref{th1}; the proof makes use of  the celebrated  Pohozaev
identity (see \cite{Po}).\\

\noindent{\it Proof of Theorem \ref{th1}}\\

Multiply (2.5) by $x\cdot \nabla w$ and $w$
respectively, and integrate the resulting equations over $D$. After some
manipulation (details can be found in \cite{Po}), one obtains the following $''$ Pohozaev-like $''$ identity\\

\begin{align}\label{po}
&\int_{D}K(\frac{1}{2}-\frac{3}{q+1})w^{q+1}dx+\int_{D} [\frac{1}{2}e^{\alpha S}|\nabla w|^2+\frac{\alpha Ke^{-\alpha S}}{q+1}w^{q+1}] (x\cdot\nabla S)dx\notag\\
&+\int_{D}(\eta A'(\eta)+\frac{5A(\eta)}{2})wdx
+\int_{\partial D}\frac{1}{2}e^{\alpha S}|\nabla w|^2(x\cdot \nu)dS=0,
\end{align}\\

\noindent where
$A(\eta)=2\Omega(\eta)(\Omega(\eta)+\eta\Omega'(\eta)$, and $\nu$
is the unit outer normal vector. If $\gamma\le 6/5$ (resp.
$\gamma< 6/5$), then $q\ge 5$ (resp. $q> 5$)  and thus
$\frac{1}{2}-\frac{3}{q+1}\ge 0$ (resp.
$\frac{1}{2}-\frac{3}{q+1}> 0$) . Thus, if $\eta
A'(\eta)+\frac{5A(\eta)}{2}>0$ (resp. $\eta
A'(\eta)+\frac{5A(\eta)}{2}\ge 0$),  $x\cdot \nu\ge 0$ and $x\cdot
\nabla S\ge 0$, (\ref{po}) implies $w(x)=0$ for $x\in D$. This
completes the proof of Theorem \ref{th1}.

\section{Further Discussion.}

In this section, we give further discussions on the difference
between our  formulation of the problem and that in [2] and [11].
As mentioned in Section 1,   for the isentropic case, i. e.,
$S=constant$,
  system (1.3) is considered in [2] and [11] in  all of $R^3$ space, with the prescribed
  total mass M. In
[2]and [11], the function $\Phi$ is given by
\begin{equation}\label{6.1}
\Phi(x)=-G\int_{R^3}\frac{\rho(y)}{|x-y|}dy.
\end{equation}
This choice is  natural  when one considers the following equation
$\Delta \Phi=4\pi G\rho$  in all of $R^3$, because any bounded
solution  $\Phi$ for this equation in $R^3$ differs from
(\ref{6.1}) only by a constant.

In our formulation, we consider system (1.3) in a bounded domain
$D$, the domain of a star. The boundary condition in our
formulation is $\rho=0$, to match the exterior vacuum
continuously. In this formulation, the function $\Phi(x)$ could be
different from the following formula
\begin{equation}\label{6.2}
\Phi(x)=-G\int_{D}\frac{\rho(y)}{|x-y|}dy,
\end{equation}
due to the boundary effects. We illustrate this by considering the
isentropic case, i. e., $S=constant$ (without loss of generality,
$S$ is assumed to be zero). In this case, the first equation in
(1.3) reduces to
\begin{equation}\label{6.3}
\nabla (w+\Phi-J)=0,
\end{equation}
where $w$ is given by (2.3) (with $S=0$), $J$ is given by (1.4).
From ( 6.3),  we have
\begin{equation}\label{6.4}
w+\Phi-J=C,
\end{equation}
where $C$ is a constant. On the boundary $\partial D$, since
$\gamma>1$, the boundary condition $\rho=0$ implies $w=0$  on the
boundary $\partial D$. This, together with (6.4),  implies that
\begin{equation}\label{6.5}
\Phi(x)=C+J, \qquad {\rm for~} x\in \partial D.
\end{equation}
The solution of the problem $\Delta \Phi= 4\pi G \rho$ with the
boundary condition (6.5) could be different form  formula (6.2).
The difference  is a harmonic function which counts  the boundary
effects.

When the domain $D$ is ball and the entropy $S$ is a constant, we
give the existence results and study some physical properties of
the solutions. Historically, incompressible fluid model of
rotating stars are studied extensively. For the incompressible
fluid model,  density is a positive constant inside a star and
zero outside the star, so there is a discontinuity of density
across the boundary of a star.  The problem for the incompressible
fluid model of Newtonian rotating stars is to determine the
boundaries of stars with the prescribed angular velocity (or
angular momentum) and total mass. For such problems, explicit
solutions are found (e. g. Maclaurin spheroids and Jacobi
ellipsoids). Those spheroids and ellipsoids are close to balls if
the angular velocities are small. For the compressible fluid
model, the problem is quite different since the distribution of
density must be determined. For the compressible fluid model of a
rotating star, if the domain $D$ is not a ball, it seems  very
difficult to obtain some interesting properties of solutions like
what we obtain in Theorems 2.1-2.3 for the case  when $D$ is  a
ball. When $D$  slightly differs from a ball, for example, an
ellipsoid with eccentricity close to one, it is
 reasonable to expect  the solutions obtained in this paper for the case when $D$ is
a ball give  good approximations.

Some non-existence results are proved in [11] for large angular
velocity $\Omega$ for the problem formulated in [2]. This means,
in order to ensure  existence of solutions for the problem
formulated in [2]; the angular velocity cannot exceed a critical
value. This critical value is given in term of total mass $M$,
which is prescribed in the formulation of the problem in [2] and
[11].  Our existence theorem, Theorem 2.1, is valid for any
angular velocity $\Omega$. This difference is due to the different
formulations of problems mentioned above. In our formulation, we
do not prescribe the total mass, instead we prescribe the domain.
The total mass thus depends on the angular velocity and the
domain. We also give a lower bound of the total mass (cf. (2.22))
in terms of angular velocity.

{\bf Acknowledgement:} Part of this work was completed when TAO
LUO was an assistant professor at the University of Michigan. JOEL
SMOLLER was supported  in part by the NSF, contract number
DMS-010-3998. We are grateful to the referee for his very
interesting remarks and comments, by which a new section, section
6 is added in the final version of the paper.

\bibliographystyle{plain}

\newpage

Tao Luo\\
Department of Mathematics,\\ Georgetown University,\\
Washington DC 20057-1233.\\
USA\\
E-mail: tl48@georgetown.edu\\

Joel Smoller\\
 Department of Mathematics,\\ University of
Michigan\\ 525 East University Ave. \\Ann Arbor, MI 48109-1109. \\
USA\\
E-mail: smoller@umich.edu\\

\end{document}